\documentclass{emulateapj}

\def\gtsim{
\mathrel{\raise.3ex\hbox{$>$}\mkern-14mu\lower0.6ex\hbox{$\sim$}}
}
\def\ltsim{
\mathrel{\raise.3ex\hbox{$<$}\mkern-14mu\lower0.6ex\hbox{$\sim$}}
}
\def\farcs{\hbox{$.\!\!^{\prime\prime}$}}
\def\deg{\hbox{$^\circ$}}

\slugcomment{Submitted to the Astrophysical Journal }

\shorttitle{O-Rich Outer Ejecta in Cas A}
\shortauthors{Fesen et al. }
\received{2005 July 30}
\begin{document}

\title{Discovery of Outlying, High-Velocity Oxygen-Rich Ejecta in Cassiopeia A\altaffilmark{1} } 

\author{Robert A.\ Fesen\altaffilmark{2}, 
        Molly C. Hammell\altaffilmark{2}, 
        Jon Morse\altaffilmark{3},
        Roger A.\ Chevalier\altaffilmark{4},
        Kazimierz J.\ Borkowski\altaffilmark{5}, 
        Michael A.\ Dopita\altaffilmark{6},
        Christopher L.\ Gerardy\altaffilmark{7},
        Stephen S.\ Lawrence\altaffilmark{8}, 
        John C.\ Raymond\altaffilmark{9}, \& 
        Sidney van den Bergh\altaffilmark{10}
                                            } 
\altaffiltext{1}{Based on observations with the NASA/ESA Hubble Space Telescope,
obtained at the Space Telescope Science Institute,
which is operated by the Association of Universities for Research in
Astronomy, Inc.\  under NASA contract No.\ NAS5-26555.}
\altaffiltext{2}{6127 Wilder Lab, Department of Physics \& Astronomy, Dartmouth College, Hanover, NH 03755} 
\altaffiltext{3}{Department of Physics and Astronomy, Arizona State University, Box 871504, Tempe, AZ 85287}
\altaffiltext{4}{Department of Astronomy, University of Virginia, P.O. Box 3818, Charlottesville, VA 22903  }
\altaffiltext{5}{Department of Physics, North Carolina State University, Raleigh, NC 27695 }
\altaffiltext{6}{Research School of Astronomy and Astrophysics, The Australian National University, 
                 Cotter Road, Weston Creek, ACT 2611 Australia}
\altaffiltext{7}{Astrophysics Group, Imperial College London, Blackett Laboratory, Prince Consort Road, London SW7 2BZ}
\altaffiltext{8}{Department of Physics and Astronomy, Hofstra University, Hempstead, NY 11549}
\altaffiltext{9}{Harvard-Smithsonian Center for Astrophysics, 60 Garden Street, Cambridge, MA 02138}
\altaffiltext{10}{Dominion Astrophysical Observatory, Herzberg Institute of Astrophysics, NRC of Canada,
                 5071 West Saanich Road, Victoria, BC V9E 2E7, Canada  }

\begin{abstract}

Analysis of broadband {\em HST} ACS and WFPC2 images of the young Galactic
supernova remnant Cassiopeia A reveals a far larger population of outlying,
high-velocity knots of ejecta with a broader range of chemical properties than
previously suspected. ACS filter flux ratios along with
follow-up, ground-based spectra are used to investigate some of the kinematic
and chemical properties of these outermost ejecta. 
In this paper, we concentrate on a $\simeq$1.5 sq arcmin region located
along the eastern limb of the remnant where numerous outer emission knots
are optically visible due to an interaction with a local circumstellar
cloud, thereby providing a more complete and unbiased look at the remnant's
fastest debris fragments. From a study of this region, we identify three main
classes of outer ejecta:
1) Knots dominated by [N~II] $\lambda\lambda$6548,6583 emission; 
2) Knots dominated by oxygen emission lines especially [O~II] $\lambda\lambda$7319,7330; and
3) Knots with emission line strengths similar to the [S~II] strong FMK ejecta commonly seen in the main emission shell.  
We identified a total of 69 [N~II], 40 [O~II], and 120 [S~II] knots in the small eastern limb region studied. 
Mean transverse velocities derived from observed proper motion for 63 N-rich, 37 O-rich, and 117 FMK-like knots 
identified in this region were found to be 8100, 7900, and 7600 km s$^{-1}$, respectively.

The discovery of a significant population of O-rich ejecta situated in between
the suspected N-rich outer photospheric layer and S-rich FMK-like ejecta
suggests that the Cas A progenitor's chemical layers were not completely
disrupted by the supernova explosion outside of the remnant's NE and SW high
velocity `jet' regions. In addition, we find the majority of O-rich outer ejecta
at projected locations out beyond (v = $6500 - 9000$ km s$^{-1}$) the remnant's
fastest moving Fe-rich X-ray emission material (6000 km s$^{-1}$) 
seen in {\sl Chandra} and {\sl XMM} data along the eastern limb. 
This suggests that penetration of Fe-rich
material up through the S and Si-rich mantle did not extend past the
progenitor's N or O-rich outer layers for at least this section of the remnant.   

\end{abstract}


\section{Introduction}

With an estimated explosion date no earlier than 1671$\pm1$ 
(Thorstensen, Fesen, \& van den Bergh 2001) 
and an estimated distance of 3.4$^{+0.3}_{-0.1}$ kpc \citep{Reed95}, the bright
radio source Cassiopeia A (Cas A) is currently the youngest known Galactic
remnant of a core-collapse supernova (SN).  The remnant consists of
an optical, infrared, and X-ray bright $4'$ diameter ($\simeq$4 pc) emission ring of
reverse shock heated SN debris rich in O, Si, S, Ar, Ca and Fe
\citep{ck78,ck79,Douvion99,Hughes00,Willingale02,Hwang03}.  Optically, the remnant's SN
debris are seen as condensations and filaments known collectively as
Fast-Moving Knots (FMKs), with expansion velocities between $4000 - 6000$ km
s$^{-1}$ \citep{min68,vdb71,Law95,Reed95}.

Estimates for the initial main sequence mass of the Cas~A progenitor are
between 10 and 30 M$_{\sun}$
\citep{Fabian80,Vink98,Willingale03,Laming03,Vink04}. A Wolf-Rayet star
progenitor has often been proposed \citep{Langer86,Fes87,Fes91,GS96} and a
mass-loss, ring nebula around such a progenitor is consistent with the presence
of slow-moving clumps (`QSFs', v = $0-400$ km s$^{-1}$) of N and He-rich,
pre-SN circumstellar mass loss material \citep{Pvdb71,Kvdb76,KC77,ck78}.

Along the outer periphery of the remnant's main ejecta shell, numerous small
knots of ejecta have been detected at projected positions near or beyond the
forward shock front (see \citealt{Fesen01} and references therein).  The
brightest and best known group of such high-velocity outer ejecta lies in the
northeastern `jet', which consists of three or four streams of ejecta knots
extending out $\sim2'$ from the remnant's northeast rim. Over a hundred jet knots
have been identified from ground-based images 
and these knots exhibit much higher proper motion derived transverse velocities
($7000 - 14000$ km s$^{-1}$) than the $\sim$ 6000 km s$^{-1}$ FMKs seen in the
main shell \citep{vdbD70,Fes96}.  Optical imaging and spectra of FMKs in the
jet show strong [S~II] $\lambda\lambda$6716,6731 but relatively weak or absent
[O~III] $\lambda\lambda$4959,5007 emission. A few much fainter outer FMKs have
been recently found in an apparent southwest `counterjet' spatially coincident
with faint extended X-ray emission \citep{Fesen01,Hwang04}. 

In addition to these outlying NE and SW jet FMKs, some 50 faint outlying
knots with strong [N~II] $\lambda\lambda$6548,6583 emission but weak or absent
H$\alpha$, [S~II] $\lambda\lambda6716,6731$ and [O II]
$\lambda\lambda7319,7330$ line emissions have been detected around
much of the remnant's periphery \citep{Fes87,Fesen01}.  Radial velocity
measurements combined with estimated transverse velocities based on proper
motion or knot displacement from the remnant's center of expansion indicate
these strong [N~II] emission knots have space velocities of $8000 - 12000$ km
s$^{-1}$ \citep{Fesen01}. These apparently N-rich knots appear to represent
fragments of the progenitor's outer layers and are
consistent with the suggestion of a WN Wolf-Rayet progenitor which experienced
substantial pre-SN mass-loss (the N-rich QSFs) before exploding as a Type Ib/c
or Type~II supernova. 

In this paper, we report results from a new and much deeper survey of outer
ejecta fragments in the Cas A SNR. The survey reveals a far larger population of
high-velocity knots of ejecta and a broader range of chemical properties than
previously suspected. A complete catalog of outer ejecta knots is beyond
the scope of the present work and will be addressed in a separate paper
\citep{HF06}. Here we concentrate on the remnant's outer ejecta rich eastern
limb region and use this area to help define some of the kinematic and chemical
properties of the remnant's fastest moving ejecta.  The observations are
described in $\S$2, with the results and conclusions given in $\S$3 and $\S$4,
respectively. 

\section{Observations}

\subsection{Imaging}

As part of an imaging survey of the ejecta in the Cas~A SNR, we
obtained high resolution images of the remnant
between January 2000 and December 2004 using two different cameras on board the
{\it Hubble Space Telescope} ({\sl HST}).

Multiband images taken at six pointings covering the entire remnant, including
all previously known outlying ejecta knots, were obtained
on 2004 March 4--6 and 2004 December 4--5 using the Wide Field Channel (WFC) of the
Advanced Camera for Surveys (ACS; \citealt{Ford98,Pavlovsky04}) on board {\sl
HST}.  The ACS/WFC consists of two $2048 \times 4096$ CCDs providing a field of
view $202'' \times 202''$ with an average pixel size of $0\farcs05$.  Four
2-point line dithered images were taken in each of the four ACS/WFC Sloan
Digital Sky Survey (SDSS) filters, namely F450W, F625W, F775W, and F850LP
(i.e., SDSS g,r,i, and z), at each target position to permit cosmic ray
removal, coverage of the $2\farcs5$ interchip gap, and to minimize saturation
effects of bright stars in the target fields.

Total integration times in the F450W, F625W, F775W, and F850LP filters were
2000 s, 2400 s, 2000 s, and 2000 s respectively.  Standard ACS pipeline
IRAF/STSDAS\footnote{IRAF is distributed by the National Optical Astronomy
Observatories, which is operated by the Association of Universities for
Research in Astronomy, Inc.\ (AURA) under cooperative agreement with the
National Science Foundation. The Space Telescope Science Data Analysis System
(STSDAS) is distributed by the Space Telescope Science Institute.} data
reduction was done including debiasing, flat-fielding, geometric distortion
corrections, photometric calibrations, and cosmic ray and hot pixel removal.
The STSDAS {\it drizzle} task was used to combine exposures in each filter.

Detected counts in each of the drizzled filter images were converted to flux
units by summing the signal in $5 \times 5$ pixel windows, subtracting a local
mean background and then multiplying by the mean flux density per unit
wavelength generating 1 count s$^{-1}$ (i.e., the `PHOTFLAM' factor) times the
filter effective bandwidth (EBW).  For the F625W, F775W, and F850LP filters,
the PHOTFLAM values used in units of 10$^{-19}$ erg cm$^{-2}$ s$^{-1}$
\AA$^{-1}$ and the EBW values used were 1.195, 1.007, 1.507, and 415.5 \AA,
434.6 \AA, and 539.4 \AA, respectively \citep{Sirianni05}.

Due to significant reddening toward Cas A (A$_{\rm V}$ = $4.5 - 8$ mag;
\citealt{Hur96,Reynoso02}), [O~III] $\lambda\lambda$4959,5007 line
emission was too weak to be detected for most outlying knots. We have,
therefore, not included F450W images in our analysis. Table 1 lists the primary
emission lines detected by the three other filters, the total system throughput
of the telescope + ACS/WFC camera + filter, and the resulting relative line
flux of a typical main shell Cas~A FMK ejecta knot.  

Previous {\sl HST} images of the remnant taken on 2000 January 23 and 2002
January 16 using the Wide Field Planetary Camera 2 (WFPC2) were also analyzed,
primarily to measure knot proper motions.  While these WFPC2 images mainly 
targeted Cas A's bright main shell \citep{Fesenetal01,Morse04}, they also 
included $4 \times 500$ s F675W filter exposures of portions of the remnant's
outer eastern limb. The WFPC2 images have an image scale of $0\farcs1$
pixel$^{-1}$ which under-sampled {\sl HST's} $0\farcs046$ angular resolution.
The F675W filter (bandpass: $6000 - 7600$ \AA) was sensitive to line emissions
of [O~I] $\lambda\lambda$6300,6364, [N~II] $\lambda\lambda$6548,6583, [S~II]
$\lambda\lambda$6716,6731, [Ar III] $\lambda$7136, and  [O~II]
$\lambda\lambda$7319,7330. Further descriptions of these data and their
reduction can be found in \citet{Fesenetal01} and \citet{Morse04}.

\subsection{Spectroscopy}
                           
Follow-up, low-dispersion optical spectra of a few outlying emission knots were
obtained on three nights in September 2004 using the 
MDM\footnote{Formerly known as the Michigan--Dartmouth--MIT Observatory.} 2.4~m
telescope and the Modular Spectrograph with a 600 lines mm$^{-1}$ 6000 \AA \
blaze grating and a $1\farcs 5 \times 4.0'$ slit. Typically, we obtained two 1000 s, 1500
s, or 2000 s exposures for each emission knot yielding a spectrum with
an effective coverage of 6000 -- 8000 \AA\ and a spectral resolution of
$\simeq$ 2 \AA.  Standard IRAF data reduction software was used with
wavelengths calibrated with Hg, Ne, and Xe lamps and \citep{Massey90} standard
stars.  Although all three nights were photometric, seeing was 
$1\farcs 0 - 1\farcs 5$ leading to considerable slit light losses at times due
to variable seeing conditions and guiding errors.  
                                                                                                                          
These data supplemented previous low-dispersion spectra of ground-based
detected outlying emission knots obtained in October 1996 using the MDM 2.4~m
telescope with the Mark III Spectrograph and a 1024 $\times$ 1024 Tektronix CCD
detector \citep{Fesen01}.  A $1\farcs 2 \times 4.5'$ slit and a 300 lines
mm$^{-1}$ $5400$ \AA \ blaze grism were used to obtain sets of two or three
$1000 - 1200$ s exposures spanning the spectral region $4500 - 7400$ \AA\ with
a spectral resolution of $\simeq$ 8 \AA. In both these and the September 2004
observations, aperture slits were oriented North-South.

\section{Results}

A comparison of the 2000 and 2002 WFPC2 + F675W filter images of a small
portion of the remnant's easternmost limb revealed about 100 high proper motion
ejecta knots. Most of these knots went undetected in earlier ground-based
imaging surveys which had concentrated on detecting strong [N~II]
$\lambda\lambda$6548,6583 emission knots \citep{Fesen01}.

Figure 1 shows the dramatic increase in the number of detected outlying ejecta
knots using broadband {\sl HST} images compared to narrowband ground-based
images.  The upper left-hand panel of the figure shows a $\simeq 75'' \times
85''$ portion of the remnant's eastern limb as seen in a 2.4~m MDM image taken
using a 6568 \AA \ filter (FWHM = 90 \AA) sensitive to ejecta knots with strong
[N~II] emission within a limited radial velocity range ($\pm 2500$ km s$^{-1}$). 
Only a few outer knots were detected in this image
\citep{Fesen01}.  The same region imaged in January 2002 using the WFPC2 +
F675W filter is shown in the lower left-hand panel of Figure 1. In contrast to
the small handful of knots detected from the ground, over 100 small clumps of
ejecta can be seen in the WFPC2 image with knot diameters near the image
resolution ($0\farcs1$ pixel$^{-1}$).  Outlying knots visible on the WFPC2
image and confirmed by the detection of proper motions $\geq$ $0\farcs75$
between the 2000 and 2002 WFPC2 images, corresponding to V$_{\rm trans}$ $\geq$
6000 km s$^{-1}$ for d = 3.4 kpc, are marked with circles on the 2002 WFPC2
image shown in the upper right-hand panel.

The location of these outlying optical ejecta knots with respect to the
remnant's outermost X-ray emission, i.e., associated with the forward shock
\citep{Gotthelf01}, can be seen by comparing the knots' positions with the
extent of the remnant's X-ray emission shown in the lower right-hand panel image of
Figure 1. This image shows a section of the 1 Msec Chandra ACIS image (epoch
2004.3; \citealt{Hwang04}) covering the same eastern region as the other panels
in the figure.  Nearly all the outlying optical knots seen here lie (in
projection) beyond the remnant's bright main emission shell and either close to
or eastward (i.e., outside) of the remnant's $\simeq$ 6000 km s$^{-1}$ forward shock
front as determined by the remnant's easternmost X-ray emission \citep{DR04}. 

\subsection{Outer Ejecta Detection and Spectral Discrimination }

The detection of so many new outlying ejecta knots in such a relatively small
area suggested that scores of undiscovered fast-moving knots of SN debris might
exist elsewhere around the whole remnant. Unfortunately, the broad sensitivity
of the WFPC2 F675W filter to a variety of elemental emission lines including
several oxygen, nitrogen, sulfur, and argon lines precluded an easy
discrimination of knot spectral properties.  On the other hand, the ACS/WFC
SDSS filters F625W, F775W, and F850LP could be used to distinguish outlying
ejecta knots having either strong [N~II] or [O~II] emissions from the remnant's
more commonly seen O + S bright FMK ejecta knots.  

For example, as shown in Table 1, while the SDSS
F625W filter, like the WFPC2 F675W filter, is sensitive to emissions from
oxygen, nitrogen, and sulfur, the SDSS F775W and F850LP filters are mainly
sensitive to just oxygen and sulfur emissions, respectively.
Using the throughputs of the ACS/WFC + SDSS filter combinations at
the wavelengths for several strong emission lines seen in Cas~A ejecta, we 
list in Table 1 observed (uncorrected for reddening) relative fluxes
for a typical S,O,Ar main shell knot (`FMK2'; \citealt{Hur96}) and the
resulting relative fluxes that would be detected using ACS/WFC with the SDSS
filters. 

The ability of these three SDSS filters to distinguish
different types of emission knots is shown in Table 2 where we list the relative
line fluxes for five well-studied main shell ejecta knots (`FMK 1- FMK 5';
\citealt{Hur96}) along with three relatively bright outer ejecta knots.
Spectra for these three outlying knots covering the $6000 - 7500$ \AA\ wavelength
region can be seen in the left-hand panels of Figure 2.  
As seen from this table, SDSS filter flux ratios for Cas~A's main
shell FMKs typically lie between $\sim0.25 - 1.0$. The one exception is FMK 4
whose observed $4000 - 10500$ \AA\ spectrum is dominated, not by [S~III]
$\lambda\lambda$9069,9531 as is usually the case, but by relatively strong [O~II]
$\lambda\lambda$7319,7330 line emission resulting in unusually high
F775W/F850LP ratio and low F625W/F775W and F625W/(F775W + F850LP) ratio 
(also called F1/(F2 + F3); see Table 3). 

In contrast to these bright main shell knots, outlying ejecta knots
can show significantly different SDSS filter flux ratios.  For example, 
Knot 15 located well above the remnant's northern limb whose $5000 - 7500$
\AA\ spectrum shows only [N~II] $\lambda\lambda$6548,6583 emission 
(V$_{\rm r}$ = +4500 km s$^{-1}$; \citealt{Fesen01})
not surprisingly exhibits F625W/F775W and F625W/F850LP values more than
an order of magnitude larger than for the five main shell ejecta knots listed.  
In similar fashion, the outer southeastern Knot 17, which shows an 6000 - 7500 \AA\
spectrum with only [O~I] and [O~II] line emissions (see Fig.\ 2), exhibits an
O/S sensitive F775W/F850LP ratio nearly three times greater than even the very
strong [O~II] $\lambda\lambda$7319,7330 emission main shell knot FMK 4.
Lastly, the outer western limb Knot 19, which shows a fairly common FMK-like spectrum
except for the addition of strong [N~II] $\lambda\lambda$6548,6583 line emission
which actually dominates the 6000 - 7500 \AA\ region (Fig.\ 2), has 2--4 times higher 
F625W/F775W and F625W/(F775W + F850LP) ratios, half the F775W/F850LP ratio, and
nearly 2--3 times the F625W/F850LP ratio of typical main shell ejecta.   

\subsubsection{ACS/WFC Color Composite Images}

{\sl HST} ACS/WFC images covering the whole remnant using the
SDSS filters were taken in March and December $2004$ in order to obtain a more
complete picture of both the number of outlying debris knots and the diversity
of their emission properties.

A qualitative look at spectral emission variations in Cas~A's high-velocity
outermost ejecta can be seen in RGB color composite images using the F625W (=
red), F775W (= green), and F850LP (= blue) image frames.  Figure 3 shows $20''
\times 20''$ sections of the individual ACS/WFC images covering only a portion
of the eastern limb shown above in Figure 1. The three greyscale images were
produced by subtracting the March from the December 2004 ACS/WFC F625W, F775W,
and F850LP filter images.  Also shown in Figure 3 (bottom right) is a three
color composite image made from the March 2004 filter data. Ejecta knots
identified through their high proper motions are indicated on each of the
frames, with previously (ground-based) identified knots (\citealt{Fesen01})
marked by knot ID numbers.

From Figure 3, one sees that many knots are only visible on one or two
individual filter images, leading to the strong color differences apparent in
the color composite image. Similar emission variations were found for ejecta
knots located in most other regions around the remnant's periphery.

Such visual inspections of the SDSS filter data set indicated not only a far
larger population of high-velocity ejecta knots than previously realized, but
also a much broader range of emission line properties over the $6000 - 10500$
\AA\ wavelength range.  This, in turn, suggested a greater chemical diversity
than just the notion of mostly sulfur-rich FMK-like ejecta in the main shell
and jet regions and nitrogen-rich knots elsewhere.
Published spectra and our new ground-based spectra of a
handful of eastern limb knots support this conclusion. 

However, both their faintness and the small angular
separations between many individual outer knots (often less than $1''$;
see Fig. 3) prevented us from obtaining ground-based spectra of many individual 
outer ejecta knots.  We have therefore used the SDSS filter images to
classify ejecta knot emission properties into broad categories consistent with
available knot spectra.

\subsubsection{SDSS Filter Flux Ratios}

In order to translate observed color composite differences into quantitative
spectral emission differences, we measured ACS/WFC fluxes from the three SDSS
filter image sets for all outer knots located along a portion of the remnant's
eastern limb using the automated source extraction software package
SExtractor \citep{Bertin96}.  In cases where the SExtractor program failed to
return a reasonable flux, or failed to return a flux at all, the knot fluxes
were calculated by hand.  In all cases, the fluxes were calculated using 5
pixel apertures.  Background estimates were performed, by SExtractor, using a
$24$ pixel rectangular annulus about the isophotal limits of the object.  
When fluxes were calculated manually,
background estimation was performed by calculating the total $5$ pixel aperture
flux in at least five positions near the object (avoiding other sources) and then
subtracting the mean computed ``background'' flux from the total object pixel
sum.  Most knots whose fluxes required manual computation were located near a
bright background source or very close to another ejecta knot.

Flux estimates were best for the brightest, well-resolved knots (above $5
\times 10^{-16}$ erg~ cm$^{-2}$ s$^{-1}$  \AA$^{-1}$), where errors are on the
$5 - 10 \%$ level.  Moderately bright knots ($5 - 50 \times 10^{-17}$ erg
cm$^{-2}$ s$^{-1}$ \AA$^{-1}$) suffered more from local background variations
and uncertainties in knot positions, resulting in flux errors near $20 - 30\%$.
Error estimates for faint knots ($1 - 50 \times 10^{-18}$ erg cm$^{-2}$
s$^{-1}$ \AA$^{-1}$) lie approximately at the $50\%$ level.  Fainter knots, or
non-detections in one of the filters, were set to the standard deviation of the
background levels of each filter as ($\sigma _{back}$ (F625W) = $3.9\times
10^{-19}$ erg cm$^{-2}$ s$^{-1}$  \AA $^{-1}$, $\sigma _{back}$ (F775W) =
$3.7\times 10^{-19}$ erg~ cm$^{-2}$ s$^{-1}$  \AA $^{-1}$, $\sigma _{back}$
(F850W) = $5.7\times 10^{-19}$ erg~ cm$^{-2}$ s$^{-1}$  \AA $^{-1}$) for the
purposes of determining knot type by dominant emission features.  

\subsection{The Eastern Limb}
                                                                                                                              
For defining the kinematic and chemical properties of the
remnant's outlying ejecta, we chose to focus on the remnant's eastern limb.  This
area exhibits an unusually high number density of knots due to an apparent
interaction of the remnant's higher-velocity ejecta with a local circumstellar
cloud \citep{Fes87}. Therefore, this region might provide a more complete and
unbiased assessment of the remnant's fast SN debris fragments. 

The region selected lies along Cas~A's eastern limb region between position
angles 78.2 and 140.6 degrees covering $\simeq$1.5 sq arcmin in size. This
avoided the NE jet to the north, which would complicate an analysis of
knot emission properties with respect to radial distance, and ends near the
southern edge of the faint circumstellar cloud seen in this region.
 
Fluxes from the F625W, F775W, and F850LP filter images were measured for a
total of 229 outer emission knots identified in this east region through high
proper motions measured using the March and December 2004 ACS/WFC images. The
resulting F625W/F775W and F775W/F850LP flux ratios for these 229 knots are
plotted in Figure 4.  Also shown in this plot are the SDSS filter ratios for
the five main shell knots (the purple star symbols) listed in Table 2, and the
three outer ejecta knots (knots 15, 17, and 19; circled numbers) discussed
above in $\S$3.1 and also listed in Table 2. 

\subsection{Emission Classes of High-Velocity Ejecta Knots}

We binned the outlying ejecta knots along the remnant's eastern limb into three
emission classes, namely, [N~II] strong knots, [O~II] strong knots, and [S~II]
strong FMK-like knots. They are color-coded in Figure 4 as red, green and blue,
respectively, consistent with their colored appearance on our composite images.
Flux ratio criteria between these three classes were chosen to segregate knots with
similar ratios seen for main shell or the outer ejecta knots with existing spectroscopy 
listed in Table 2. 

Specifically, we chose a flux ratio for F775W/(F625W + F850LP) $\geq$ 1.0 to
separate out the [O~II] strong FMKs from the [S~II] strong FMKs; that is, those
knots with stronger [O~II] $\lambda\lambda$7319,7330 emission detected via the
F775W filter than the combined strength of [N~II], [S~III] and [S~II] emissions
detected in the F625W and F850LP filters.  Similarly, knots with strong [N~II]
emissions were selected via F625W/(F775W + F850LP) $\geq$ 1.0, thereby
selecting those knots where the combined flux of [O~I]
$\lambda\lambda$6300,6364, [S~II] $\lambda\lambda$6716,6731, and [N~II]
$\lambda\lambda$6548,6583 emissions was greater than the sum of F775W flux, due
mostly to [O~II], and F850LP flux sensitive to the near-infrared [S~III] and
[S~II] emissions. This latter ratio division is shown as a horizontal line in Figure 4.
Since the observed [O~I] flux rarely, if ever, exceeds the
[O~II] $\lambda\lambda$7319,7330 flux, and the observed [S~II]
$\lambda\lambda$6716,6731 emission is unlikely to ever exceed the combined flux
of [S~III] $\lambda\lambda$9069,9531 and [S~II] $\lambda\lambda$10287--10370
line emissions \citep{Hur96,Winkler91}, then any knot for which the
F625W/(F775W + F850LP) $\geq$ 1.0 requires the presence of significant [N~II]
$\lambda\lambda$6548,6583 emission.

As shown in Figure 4, the chosen ratio criteria successfully grouped the outer
FMK-like knots in with typical main shell FMKs (`FMK 1-3' and `FMK~5'; shown as
purple stars), the strong oxygen emission line ejecta like Knot 17 and FMK 4
into an [O~II] bright knot class, and ejecta with dominant [N~II] emission like
Knots 15 and 19 into a [N~II] knot category.

Measured F625W, F775W and F850LP fluxes for our selected eastern limb region
placed 69 knots in the upper `[N~II] knot' portion of Figure 4, 40 `[O~II]
knots' in the lower right-hand portion, and 120 `FMK-like knots' in the center
portion of the figure.  An arrow on a knot's plotted position indicates an
upper or lower flux limit due to a non-detection for one or more filters (i.e.,
one sigma of the background flux limit, see $\S$3.1.2).  For instance,
knots with arrows pointing to the left-hand portion of the plot, like some
[N~II] knots, have no detectable F775W ([O~II]) emission above background sky
variations. 

A knot marked on Figure 4 with a crossbar means it was undetected in one of the
filters making up its y-axis value. Since the y-axis is a combination of filter
fluxes (y = F1/(F2 + F3)) if there was no detected flux in one
filter, say filter F2, one could not simply mark the knot with a directional
arrow on the y-axis.  However, the non-detection does lead to an uncertainty in
y-axis value for that knot and the crossbar is meant to reflect that
uncertainty. The length of the crossbar varies across the plot because: i) this
is a log plot so a constant value has a different length along the axis, and
ii) more importantly, the effect of an uncertainty of say F2 on the y-axis
value depends on the specific values of F1 and F3. Crossbars were used instead
of arrows because we wished to emphasize that a non-detection in a filter would
not move a knot's plotted position into another knot category (below the line
at y = 1) no matter what the actual value of flux in the missing filter turned
out to be.

The line of double arrowed [N~II] knots appearing near the top of the [N~II]
section of Figure 4 represent eastern limb knots that have F625W/F775W  ratios
$\geq$ 20 and were undetected in both F775W and F850LP frames.  These knots
appear to represent highly [N~II] emission dominated ejecta, even more so than
the outer Knot 15 whose spectrum shows no [S~II] $\lambda\lambda$6716,6731 or
[O~II] $\lambda\lambda$7319,7330 emissions (see Fig.\ 2). Knots near the upper
part of the [N~II] strong knot section of Figure 4 have F625W/(F775W + F850LP)
ratios approaching 50 or more.

Below we briefly discuss three main types of high-velocity, outer
emission knots found in this ACS/WFC survey.

\subsubsection{[N~II] Emission Knots}

Within our selected eastern limb region, dozens of knots were found to show a
F625W flux comparable or greater than the total F775W + F850LP fluxes. For
these cases, the F625W flux is most probably due to strong [N~II]
$\lambda\lambda$6548,6583 emission rather than H$\alpha$, [S~II], or [O~I]
emissions.  This is because of the 43 strong [N~II] emission outer knots with
published line strengths [O~I] and [S~II] emissions were not detected and
[N~II] emission was always much stronger than H$\alpha$, with only three
exhibiting H$\alpha$ emission at readily detectable levels, i.e., [N~II]
$\lambda$6583/H$\alpha$ =  3 -- 4 \citep{Fesen01}.  Thirty of the 43 knots were
bright enough to set meaningful [N~II] $\lambda$6583/H$\alpha$ lower limits and
had typical values $\geq$5, with 20\% above 10.
 
If the observed F625W emission was due to [O~I] $\lambda\lambda$6300,6364
and/or [S~II] $\lambda\lambda$6716,6731 emission then, based on existing
spectra, these knots should have been easily detected through [O~II] emission
in the F775W images or through strong [S~II] and [S~III] emission in F850LP
images.  This was not the case for 69 eastern limb knots indicating these knots
have an optical ($6000 - 10500$ \AA) spectrum with a significant contribution
from, if not dominated by, [N~II] emission. 

Many of the brighter F625W strong emission knots match known [N~II] dominant
emission knots previously identified through ground-based images and spectra
(e.g., Knots 4, 5, and 7). A $6000 - 7500$ \AA\ spectrum of an eastern limb
ejecta clump not previously studied spectroscopically, labeled `East 1', is
shown in the upper right-hand panel of Figure 2 (see Table 3 for coordinates
and fluxes). 

A spectrum of another eastern limb knot, `East 3', with F625W emission due in
part to the presence of moderately strong [N~II] emission is shown in the lower
right-hand panel of Figure 2.  This knot failed our strong [N~II] knot flux
criteria test (i.e., F625W/(F775W + F850LP) $\geq$ 1.0) and was classified as
an FMK-like outer knot even though, unlike any main shell FMK, it shows some
appreciable [N~II] emission.

\subsubsection{[O~II] Emission Knots}

Both our color composite images and the plotted knot filter fluxes on the right-hand
portion of Figure 4 indicated a surprising number of knots with spectra
dominated by F775W flux.  Strong F775W emission indicates a high probability
for strong [O~II] $\lambda\lambda$7319,7330 line emission since the only other
significant emission lines in the F775W filter's passband are the [Ar~III]
$\lambda\lambda$7136,7751 lines.  However, even for those rare knots having
exceptionally strong argon emission lines  (e.g., the `calcium FMK';
\citealt{Hur96}), [S~II] and [S~III] emissions still dominate the optical spectra
and thus would be plotted on Figure 4 in with the more common [S~II] strong
FMKs. Therefore, strong F775W emission is a good marker
for unusually strong, if not dominate, [O~II] line emission.   

In the eastern limb region, 40 knots were identified as having a spectrum
dominated by [O~II] emission.  Examples of these [O~II] F775W bright knots can
be seen in the color composite image of Figure 3 (lower right-hand panel).  The
group of green knots in the lower right portion of the color composite make up Knot
17A, an ejecta clump first identified from ground-based images and exhibiting
spectra similar to the more southern Knot 17 (see Fig.\ 2; middle panels).  This
knot was one of a small cluster of outer ejecta knots (Knots 16, 17, 17A, and
17B) for which earlier low-dispersion spectra showed strong oxygen emission lines
\citep{Fesen01}.  Their limited spatial distribution along a section of the SE
limb had initially made them appear to be simply a few odd, high-velocity ejecta clumps, not
representatives of a large and well distributed subclass of outer ejecta which
we now identify as strong [O~II] outer knots.

\subsubsection{FMK-like Knots}

In our color composite images, FMK-like knots appear white or purple due to
strong F625W and F850LP fluxes arising principally from [S~II]
$\lambda\lambda$6716,6731 and [S~III] $\lambda\lambda$9069,9531 emissions.
These knots typically occupy the middle of the filter flux ratio plot shown in
Figure 4, as do the four bright main shell knots previously discussed and shown
here as purple star symbols.

Earlier ground-based surveys had found that nearly all outlying FMK-like knots
were located in either the northeast or southwest jet ejecta regions.  However,
the deeper and higher spatial resolution {\sl HST} images reveal a more
extensive distribution of such high-velocity FMK-like ejecta.  In the eastern limb
section for example, we identify 120 [S~II] strong FMK-like outer knots. Virtually all of
these knots are small and faint, and consequently, they fell below
the detection limit of the moderately deep and broadband (FWHM =
250 \AA) [S~II] $\lambda\lambda$6716,6731 imaging survey of the remnant
\citep{Fesen01}. 

\subsection{Outer Knot Kinematics and Distribution}

Histograms for the radial distances for all 229 identified knots in our
selected eastern limb region grouped by emission type are shown in Figure 5.
Radial distances for these knots, measured from the remnant's estimated center
of expansion (\citealt{Thor2001}), are in the range $135'' - 195''$ ($2.2 -
3.2$ pc).  As shown in this figure, [N~II] bright knots tend to lie at greater
distances (in projection) than the [O~II] bright knots which, in turn, lie
coincident with or slightly ahead of the FMK-like knots.  Mean distances for
the [N~II], [O~II], and FMK-like knots are 168$''$, 163$''$, and 158$''$ (= 2.8
pc, 2.7 pc, 2.6 pc), respectively.  The observed range of radial distances,
when converted to average transverse velocities (assuming an age of 320 yr and a
distance of 3.4 kpc) correspond to velocities of $7000 - 9500$ km s$^{-1}$.  

For 217 of the 229 eastern limb knots we were able to estimate transverse 
velocities directly from proper motion measurements.  Positional displacements were
measured using WCS aligned January 2002 WFPC2 F675W and March 2004 ACS/WFC F625W,
F775W, and F850LP images. Measured proper motions relative to observed
knot radial distances are plotted in Figure 6. The line shown is a least
squares fit to the radial distance vs.\ proper motion for all knots and has a
slope of $2.9 \pm0.2$ mas yr$^{-1}$ arcsec$^{-1}$. The [N~II] knots exhibit the
highest proper motions, up to $0\farcs6$ yr$^{-1}$, with somewhat smaller
systematic values for the [O~II] and FMK-like knots.  Transverse velocities
derived from these proper motions grouped by knot emission type yield mean
expansion velocities of 8100, 7900, and 7600 km s$^{-1}$ for 63 N-rich, 37
O-rich, and 117 FMK-like knots, respectively.  

Consistent with their greater average radial distances
(Fig.\ 5), [N~II] knots show the highest maximum transverse velocities, up to
9300 km s$^{-1}$, followed by the FMK-like knots with velocities up to 9100 km
s$^{-1}$, and then [O~II] knots of up to 9000 km s$^{-1}$.  These velocities
significantly exceed the maximum space velocity of 6000 km s$^{-1}$ seen for
main shell ejecta \citep{Law95,Reed95}.  However, knots in all three classes
with the smallest radial distances ($\simeq$ $140''-160''$) show a marked
decrease in observed proper motions and these tend to lie at a projected
positions just ahead of main shell knots.  This effect is seen for all three
knot emission classes and may indicate stronger deceleration by the reverse
shock. Conversely, the true slope may actually be steeper than shown with the
most distant knots the most decelerated.

Finally, the location of these outlying optical knots categorized by their
measured filter flux ratios relative to the remnant's faint outermost X-ray
emission is shown in Figure 7. Here the March 2004 red [N~II], green [O~II],
and blue FMK optical knot positions are overlaid onto the {\sl Chandra} 1 Ms
exposure (right panel), itself color coded by isolating Si (1.8--2.0 keV), Fe K
(6.5--7.0 keV), and continuum (4.2--6.3 keV) emissions following
\citet{Hwang04}. The three eastern limb knots (E1--E3) discussed above for
which we obtained spectra are marked in the left-hand panel which shows just
the remnant's Fe X-ray emission detected by {\sl Chandra}.  One sees from
this figure that nearly all the outer eastern limb knots lie near or out beyond
the remnant's forward shock front. Importantly, nearly all the strong [O~II] knots which
we have identified via filter ratios lie (in projection) out ahead of the
remnant's bright Fe-rich X-ray emitting ejecta.

\section{Discussion}

A remnant's outermost debris may carry clues about both the chemistry of the progenitor's outer
layers and about the dynamics of the SN explosion. Our {\sl HST} imaging of the Cas A
remnant has revealed a surprisingly large population of fast-moving ejecta
outside of the NE and SW jet regions.  These ejecta knots lie coincident with
or out beyond the remnant's current forward shock front position (as seen in
X-rays; see Fig.\ 1) and exhibit a wide range of filter flux ratios indicative
of a significant chemical diversity, confirmed by spectral data (Figs.\ 2--4).  

While optically emitting debris constitute only a small fraction of the total
ejected mass, the remnant's fastest moving material is perhaps best studied
optically. For example, the NE and SW jets can be optically traced about 80
arcsec farther out than in radio or X-rays and only a handful of outer ejecta
knots around the rest of the remnant are visible in even the deepest radio or
X-ray images. On the other hand, outlying ejecta knots are seen in the optical
due to their interaction with surrounding ISM/CSM which generate $\sim$100 km
s$^{-1}$ internal shocks driven into the knot by the high stagnation pressure
behind the knot's bow shock. Consequently, what one sees 
may be a biased view of the distribution of outer ejecta.
 
Because the visibility of outlying optical knots depends upon whether they have
recently encountered local ISM/CSM clouds or the forward and reverse shock
fronts, combining results from different regions around the remnant might lead
to a seemingly greater dispersion of chemical abundance variations with radial
distance.  This, plus the unlikely smooth and spherical expansion of the
progenitor's outer layers, make uncertain apparent correlations of outer ejecta
kinematic versus chemical properties when summed across the whole remnant.
Consequently, we restricted the present study to just a relatively small,
limited region along the Cas A remnant's eastern boundary where the ejecta
appear to be interacting with an extended CSM cloud.  

\subsection{Outer Knot Abundances}

General spectral properties of outer emission knots were determined 
by the use of filter flux ratios and we have identified three main classes of
outer ejecta: 1) Knots dominated by [N~II] $\lambda\lambda$6548,6583 emission;
2) Knots dominated by oxygen emission lines especially [O~II]
$\lambda\lambda$7319,7330; and 3) Knots with emission line strengths much like
those seen in the FMKs found in the main shell.  The histograms shown in Figure
5 make a compelling case that, at least for the eastern limb region studied, the
emission-line properties and hence likely chemical properties of Cas~A's
highest velocity ejecta correlate fairly well with expansion velocity.

Because few spectra are available for the nearly 230 outer knots found within
our selected eastern limb region, accurate and unambiguous abundance ranges are
not possible for the different knot types from the broadband filter
measurements.  However, the observed range of flux values, when taken together
with representative knot spectra (see Fig.\ 2), offer qualitative estimates on
relative nitrogen, oxygen, and sulfur abundances.

Using the modified \citet{Raymond79} shock code described in \citet{Blair00},
several models were run for shock speeds of 50 -- 100 km s$^{-1}$ assuming knot
preshock densities $\simeq 10^{2-3}$ cm$^{-3}$.  The log of the abundances of
oxygen and nitrogen were varied from 14 to 16 (H = 12.00) with a sulfur
abundance of 14 and other elements heavier than Ne set at or below solar
values. These model results indicate that while [S~III]/[O~II] is sensitive to shock
velocity, the ratio of [N~II] $\lambda\lambda$6548,6583/[O~II]
$\lambda\lambda$7319,7330 is fairly independent of shock speed and the
abundances of other elements.  For N/O abundance ratios of 0.1 and 10, we found
[N~II]/[O~II] values of $\simeq$ 0.5 and $\simeq$ 100, respectively. Such
values suggest that for strong [N~II] emission outer knots, i.e., knots plotted
in the uppermost portion of Figure 4 such as East 1 and Knot 15 and having
spectra with [N~II]/[O~II] $>$ 20 (cf.\ Fig.\ 2), the N/O abundance ratio
exceeds unity indicating a nitrogen overabundance at least an order of
magnitude over solar.

The highest-velocity ejecta in the east limb region are clearly those with
dominate [N~II] emission (see Figs. 5 and 6).  This fact, when taken together
with the presence of N-rich circumstellar material (QSFs) and the observed lack of
appreciable hydrogen emission in all but a couple of outermost ejecta
\citep{Fesen01} support the notion that the Cas~A progenitor was probably a WN
Wolf-Rayet star that exploded as a core-collapse SN Ib event (Fesen et al.\
1987, Fesen \& Becker 1991, Vink 2004).  Although recent estimates for the
progenitor mass based on X-ray abundance analysis have difficulty explaining
the lack of C-burning products such as Ne and Mg, the remnant's estimated
oxygen mass of 1--3 M$_{\odot}$ plus the excellent correlation between Si, S,
Ar, and Ca abundances suggest a relatively high main sequence mass $\simeq$
15--25 M$_{\odot}$ \citep{Woosley93,WW95,Willingale02} not unlike that expected
for a WR star progenitor.

For the strong [O~II] emission knots, both our shock models as well as the metal-rich
shock models presented in \citet{Morse04} which assumed somewhat lower preshock densities
suggest that a near solar abundance ratio for O/S $\sim$ 10 yields an observed
[O~II] $\lambda\lambda$7319,7330/[S~II] $\lambda\lambda$6716,6731 ratio of
around 1 -- 2 and a F775W/F850LP filter ratio of around one (after correcting
for an A$_{\rm V}$ = 5 mag and the ACS/WFC throughput). Because [O~II] strong knots
show F775W/F850LP ratios above 1.5 (and can exceed 10) and are unlikely to have
strong [S~II] $\lambda\lambda$6716,6731 emission in light of F625W/(F775W +
F850LP) ratios between 0.07 -- 0.5, such knots appear to possess O/S abundance
ratios many times above the solar value. This qualitative assessment is
consistent with spectra taken of of such knots, (e.g., knot East 2 and Knot 17;
see Fig.\ 2) for which only emission lines of oxygen are detected in the 6000
-- 7500 \AA \ region. 

\subsection{Asymmetric Element Mixing}

Our discovery of a significant population of O-rich ejecta outside the forward
shock front both here along the eastern limb region and elsewhere around the
remnant \citep{HF06} may help in understanding the degree and
asymmetry of compositional mixing of the Cas~A SN debris.  On the one hand, the
very high-velocity Si and S rich ejecta in the NE and SW jets plus the presence
of highly Fe-rich ejecta produced by explosive Si burning appearing outside the
remnant's Si-rich shell along the NW and SE limbs has been taken as evidence
for large scale turbulence and mixing due to a non-spherical expansion
\citep{Hughes00,Fesen01,Willingale02,Hwang03}.  This would appear consistent with some core-collapse
models which show significant mixing and overturning of ejecta through Rayleigh-Taylor
instabilities \citep{Kifon00,Kifon03}, and which SN~1987A gave direct evidence for in
the form of the apparent transport of freshly synthesized $^{56}$Ni from the
core to the H-rich envelope \citep{Arnett89}.

However, the presence of a layer of O-rich ejecta knots situated in between the
apparent N-rich outer photospheric layer and the S-rich, FMK-like ejecta layer
suggests that, at least for this limited eastern limb region away from the
remnant's NE and SW jets, Cas A's chemical layers do not appear
to have been completely mixed or disrupted all the way out to the surface.
The outlying O-rich knots we see in Cas~A exhibit spectral properties not unlike the
O-rich ejecta seen in the LMC/SMC remnants N132D and 1E~0102.2-7219 where there
is no evidence supporting mixing of O-rich ejecta with O-burning products,
i.e., Si, S, Ar, and Ca rich ejecta \citep{Blair00}.  This raises questions
about just how well mixed were the O-rich and S-rich layers in the Cas A SN on
both a local and global scale. 

The degree and uniformity of elemental mixing in Cas~A has been a long standing
question (Chevalier \& Kirshner 1978, 1979; Johnson \& Yahil 1984;  Winkler et
al.\ 1991; Douvion et al.\ 1999; Hwang et al.\ 2000, Willingale et al.\ 2002).  
Strong O and S lines in
most main shell ejecta knots indicate that substantial mixing of the O-rich and
Si+S-rich layers did indeed occur.  However, there is also compelling evidence that
this mixing was neither microscopic nor homogeneous on large scales.  Some
ejecta clumps have optical spectra with nearly only oxygen emission lines
visible \citep{ck79,Winkler91}, while others have very weak or nearly absent
oxygen emissions but unusually strong lines of Ca, Ar, and S \citep{Hur96}.
Compared to the range of [S~II]/[O~II] emission line ratios for optical knots in
the main shell, there is a fairly constant [Ar~III] $\lambda$7136/[S~III]
$\lambda$9069 ratio suggesting Ar and S formed at about a constant ratio from
oxygen burning, in sharp contrast to the observed [O~III] $\lambda$5007/[S~III]
$\lambda$9069 ratio which varies by as much as a factor of 100.  

The observed range of ejecta spectral properties could reflect either 
different degrees of explosive oxygen burning \citep{ck79} or varying levels 
of mixing between outer O-rich material with inner layers of O-burning products
\citep{Winkler91}. Ne and Ar line emission maps made from mid-infrared
observations show an anti-correlation between the presence of neon (which is
most abundant in the outermost O-rich layers) and silicon-rich ejecta pointing 
to a mixing process that was not homogeneous on large scales
(Douvion et al.\ 1999).

The fact that there is no correlation of main shell ejecta abundances with
expansion velocity means that there had to be some inversion of the S-rich
inner layers relative to the O-rich outer layers \citep{ck79}. Large scale
Rayleigh-Taylor instability `fingers' or incomplete explosive O-burning across
different regions might help explain the degree of mixing observed. 
In and around R-T fingers, considerable mixing would be expected 
leading to O-rich and S-rich ejecta regions with a lack of element vs.\ ejection
velocity correlation like that observed.  

\citet{Kifon00,Kifon03} found for a Type Ib SN model with 4.2 M$_{\odot}$ that
R-T instabilities form at the Si/O and O/He compositional interfaces.  This
lead to fragmentations of the Fe-rich core and considerable mixing of the
inner 2 M$_{\odot}$ which ends up with an expansion velocity as high as 3500 --
5500 km s$^{-1}$. However, this velocity is less than that seen in Cas~A and
nearly pure Fe-rich material out beyond the Si-rich layers in Cas~A's NW and SE
sections \citep{Hughes00,Willingale02,Hwang03} would seem more consistent with
a picture of strong but only regional overturning.

\cite{Kifon03} also noted that the pattern of R-T overturning carries information
about the geometry of neutrino-driven convection that seeds the R-T
instabilities. If that is true, then the observed $\sim45\deg$ angular
size of the Fe-rich areas in Cas A may be indicating the rough scale of such
neutrino-driven convection seeds. 

On the other hand, high-velocity (v = $6500 - 10000$ km s$^{-1}$)  N-rich and
O-rich ejecta at projected locations along the eastern limb out beyond the
remnant's fastest moving Fe-rich X-ray emission material (6000 km s$^{-1}$)
show that even though Fe-rich core fragments apparently penetrated the
S + Si-rich mantle ejecta layer, they did not expand past the N or O-rich outer
layers as happened in SN~1987A.  Thus in Cas~A, at some distance radially
above the R-T finger instabilities, the He + N-rich and O-rich layering structure
appears to have survived largely intact. 

A nitrogen-oxygen-iron layering can be seen in Figure 7 where the
majority of optical N-rich and O-rich knots lie in projection ahead of the
Fe-rich ejecta detected in the X-rays. While not lying exactly in the plane of
the sky, this Fe-rich material is seen to extend out to a maximum radius of
$145'' - 165''$, in this region, corresponding to an average expansion velocity of
around $7300 - 8300$ km s$^{-1}$.  De-projection of its observed radial velocity
of $-1000$ to $-1500$ km s$^{-1}$ \citep{Willingale02} would only move it a few
arcseconds farther eastward, insufficient to extend it out beyond either the
N-rich or O-rich outer ejecta knots which lie at radial distances of $150'' -
190''$ and $145'' - 185''$, respectively (see Fig.\ 5).    

\citet{Hwang03} and \citet{Laming05} comment that any Fe-rich debris ejected farther out than seen
in the {\sl Chandra} image would have encountered the reverse shock at an
earlier epoch and thus could have cooled down below the X-ray emitting
temperature range.  However, the coincidence of the inner edge of the O-rich
ejecta we have detected with the outermost Fe-rich X-ray emitting material seen
in {\sl Chandra} Fe~K image is striking (left panel, Fig.\ 7) suggesting that we're
seeing most of the fastest-moving Fe-rich material in this region. This holds 
true for the cooler Fe~L emission as well which has about the same spatial distribution and
extent as that of Fe~K \citep{Hwang00}.

However, there may be a causal relation between Fe-rich
ejecta and formation of these outlying ejecta knots. As can been seen from
Figure 7, the projected space density of the outer optical ejecta knots may be
correlated with the outermost Fe-rich X-ray emitting ejecta. For example, the
cluster of optical knots around Knot E2 lies at the easternmost tip of the
detected Fe-rich X-ray emission, with the number of knots showing a noticeable
decline to both the north and south. Without knot radial velocity information,
one cannot test if such projected correlations are in fact spatially real, yet the
coincidences seen here are intriguing.  

While radially limited layer mixing may explain the abundance pattern seen
along the remnant's east limb, this process clearly did not operate in the NE
and SW jet regions. Thus the global picture of the Cas~A SN explosion which
emerges is one of regional and radially limited overturning of mantle and core
material. Added to this scenario are the seemingly opposing NE and SW jets of
much higher velocity ejecta, rich in explosive O-burning products, whose nature
is not well understood. This all created the fairly complex SN remnant ejecta
structure exhibiting regionally diverse abundance mixing patterns we
observe today.  

\section{Conclusions}

We present an analysis of broadband {\em HST} ACS and WFPC2 images of the young
Galactic supernova remnant Cassiopeia A, and concentrate specifically on a
$\simeq$1.5 sq arcmin region located along the easternmost limb of the remnant.
This region exhibits numerous outer emission knots that are optically visible
due to an apparent interaction with a local circumstellar cloud, and therefore
provide a more complete and unbiased look at the remnant's fastest debris
fragments. Calibrated ACS filter flux ratios along with follow-up, ground-based
spectra were used to investigate some of the kinematic and chemical properties
of these outermost ejecta. These data revealed the following:

1) A substantial population of outlying, high-velocity ejecta knots exists
around the remnant which exhibit a broader range of chemical properties than
previously recognized. 
Based on the outlying knots found in the east limb region, we 
identify three main classes of outer ejecta: a) Knots dominated by [N~II]
$\lambda\lambda$6548,6583 emission, b) Knots dominated by oxygen emission lines
especially [O~II] $\lambda\lambda$7319,7330, and c) Ejecta with emission line
strengths similar to the common [S~II] strong FMK ejecta of the main shell.
Mean transverse velocities derived from observed proper motion for 63 N-rich,
37 O-rich, and 117 FMK-like knots found in this region are 8100, 7900, and 7600
km s$^{-1}$, respectively.

2) The discovery of a significant population of O-rich ejecta situated in
between the suspected N-rich outer photospheric layer and S-rich FMK-like
ejecta suggests that the Cas A progenitor's chemical layers were not completely
disrupted by the supernova explosion outside of the remnant's NE and SW high
velocity `jet' regions. 

3) We find that most O-rich outer ejecta along the remnant's eastern limb lie
at projected locations out beyond (v = $6500 - 9000$ km s$^{-1}$) the remnant's
fastest moving Fe-rich X-ray emission material (6000 km s$^{-1}$) along the
eastern limb. This suggests that penetration of Fe-rich material up through the
S and Si-rich mantle did generally not extend past the progenitor's N or
O-rich outer layers as had been previously assumed.

The chemical and dynamical picture of Cas~A that emerges is one of large,
regional overturning involving both mantle and core material, separated by
areas where the initial He + N-rich, O-rich, and Si + S-rich layering
survived fairly intact. This generated a fairly complex remnant structure with
numerous clumps of reverse shocked ejecta showing a wide diversity of
chemical abundances.
 
Future investigations  of the Cas A remnant using this {\sl HST} image data set  
will include a full catalog of all optically visible outlying ejecta knots, analysis
of the remnant's ejecta expansion asymmetries, and knot emission variability and
ablation effects due to high speed passage through the local ISM/CSM. 

\acknowledgments

We thank P. Hoeflich for helpful discussions, D. Patnaude for help
with the Chandra X-ray data reduction, and J. Thorstensen for assistance with
the celestial coordinate transformations. This work was supported by NASA through
grants GO-8281, GO-9238, GO-9890, and GO-10286 to RAF and JM from the Space Telescope Science
Institute, which is operated by the Association of Universities for Research in Astronomy.
RAC is supported by NSF grant AST-0307366 and CLG is supported through UK PPARC grant PPA/G/S/2003/00040.

\clearpage
\newpage

\begin{deluxetable}{lcclccc}
\tabletypesize{\scriptsize}
\tablecaption{HST ACS/WFC Filter Observations and Detected Line Emissions }
\tablewidth{0pt}
\tablehead{\colhead{} &  \colhead{Exposure}  &  \colhead{Filter}  &    \colhead{Main Line Emissions} & 
           \colhead{ACS/WFC+Filter} &  \colhead{Rel. Observed}
         & \colhead{ACS/WFC+Filter} \\ 
\colhead{Filter} &  \colhead{times}  & \colhead{Bandpass$^{a}$}   &  \colhead{in Filter Bandpass} & \colhead{Throughput} 
                 &  \colhead{FMK Flux$^{b}$} & \colhead{FMK Flux}   } 
\startdata
F625W  & $4 \times 600$ s     & 5450--7100 \AA  & [O I] 6300,6364         & 0.42        & ~39     & 16       \\
(SDSS r)       &   &    & [S II] 6716,6731        & 0.44        & 100     & 44       \\
               &   &    &  [N II] 6548,6583      & 0.43        & ~ ~0    & ~0       \\
               &   &   &                        &             &         &  \\
F775W & $4 \times 500$ s    & 6850--8600 \AA   & [Ar III] 7136       & 0.38        & ~41     &  15       \\
(SDSS i)  &   &   & [O II] 7319,7320        & 0.42        & 171     &  72       \\
          &   &   & [Ar III] 7751         & 0.38      & ~14     &  ~5       \\
               &      &                        &             &         &  \\
F850LP         & $4 \times 500$ s & 8500--10500 \AA     & [S III] 9069            & 0.20        & 210     & 42      \\
(SDSS z)      &   &   & [S III] 9531           & 0.14        & 600     & 85      \\
               &   &   & [S II] 10287--10370   & 0.05        & 380     & 19      \\ 
\enddata
\tablenotetext{a}{Listed wavelengths represent approximate filter transmission cut-on and cut-off points.}
\tablenotetext{b}{A Cas A FMK ejecta spectrum (`FMK 2'; \citealt{Hur96}). Values are 
                  relative to F([S II] 6716,6731) = 100.}
\end{deluxetable}

\begin{deluxetable}{lcccccccc}
\tabletypesize{\scriptsize}
\tablecaption{Observed Relative Filter Fluxes and Ratios for Cas A Ejecta Knots$^{a}$ }
\tablewidth{0pt}
\tablehead{ \colhead{Rel.\ Line Fluxes,} &
 \colhead{FMK 1} & \colhead{FMK 2}  & \colhead{FMK 3} &  \colhead{FMK 4} &  \colhead{FMK 5}  
     &  \colhead{Knot 19}   &   \colhead{Knot 15} &  \colhead{Knot 17} \\
\colhead{Filter Fluxes} &
 \colhead{(Shell)} & \colhead{(Shell)}  & \colhead{(Shell)} &  \colhead{(Shell)} &  \colhead{(Shell)}
  &  \colhead{([N II]+FMK)} &  \colhead{([N ~II] Knot)}   &   \colhead{([O II] Knot)} }
\startdata
~[O~I] ~~6300,6364   & 55   &   39  &  29    & 152   & 157   & 7   & $\leq$5  & 22         \\
~[N~II] ~6548,6583   & $<$1 &  $<$1 &  $<$1  & $<$1  &  $<$1 & 130 & 100      & $\leq$8    \\
~[S~II] ~~6716,6731  & 100  &  100  &  100   & 100   & 100   & 62  &  $\leq$5 & $\leq$8     \\
~[O~II] ~7319,7330   & 275  &  171  &  104   & 1530  & 112   & 33  & $\leq$5  & 100        \\
~[S~III] ~9069,9531  & 689  &  810  &  440   & 1190  & 558   & \nodata & \nodata & \nodata  \\
               &        &         &       &        &       &      &         &       \\
~F625W (F1)    &  ~67   & ~60     & ~56   & 108    & ~55   & 100  & 100         & 100   \\
~F775W (F2)    &  134   & ~92     & ~55   & 673    & ~58   & ~24  & ~$\leq$8    & 253  \\
~F850LP (F3)   &  135   & 146     & ~78   & 234    & 103   & ~78  & $\leq$10    & ~30  \\ 
               &        &         &       &        &       &      &         &       \\
~F625W/F775W   &  0.50  & 0.65   & 1.00   & 0.16   & 0.95  & 4.17 & $\geq$12.0  & 0.40 \\
~F775W/F850LP  &  1.02  & 0.63   & 0.71   & 2.88   & 0.56  & 0.31 & $\simeq$0.80& 8.34  \\
~F625W/F850LP  &  0.49  & 0.41   & 0.72   & 0.46   & 0.53  & 1.28 & $\geq$10.0  & 3.30 \\ 
~F1/(F2 + F3)  &  0.25  & 0.25   & 0.42   & 0.11   & 0.34  & 0.98 & $\geq$5.5   & 0.35 \\
\enddata 
\tablenotetext{a}{FMKs 1--5 are bright main shell ejecta clumps with published optical spectra \citep{Hur96}. 
                 Values are observed fluxes (relative to F([S~II] 6716,6731) = 100 where present) multiplied by
                 the ACS/WFC + filter throughputs (see Table 1). Knots 15, 17 and 19 are ejecta knots which 
                 lie outside the remnant's main shell of emission and have published optical spectra \citep{Fesen01}. 
                 Filter flux values for these knots were derived from ACS/WFC images with F(F625W) = 100.}
\end{deluxetable}
                                                                                                         
\begin{deluxetable}{ccccccc}
\tabletypesize{\scriptsize}
\tablecaption{Coordinates and Filter Fluxes for Eastern Limb Knots 1 -- 3}
\tablewidth{0pt}
\tablehead{ \colhead{Knot ID} &  \colhead{$\alpha$(J2000)$^{a}$}  & \colhead{$\delta$(J2000)$^{a}$} &  
  \colhead{F625W/F775W} & \colhead{F775W/F850LP} & \colhead{F625W/F850LP} & \colhead{F1/(F2 + F3)} } 
\startdata
 East 1   & 23:23:51.01 & 58:49:24.9 & 5.76  & 0.63  & 3.62  &  2.22   \\
 East 2   & 23:23:46.85 & 58:48:17.4 & 0.28  & 3.18  & 0.88  &  0.21   \\
 East 3   & 23:23:51.05 & 58:49:06.2 & 0.98  & 1.12  & 1.10  &  0.52   \\ 
\enddata
\tablenotetext{a}{Positions were measured for epoch 2004.3}
\end{deluxetable}

\newpage

%
%

\clearpage

\begin{figure} 
\epsscale{0.90}
\plotone{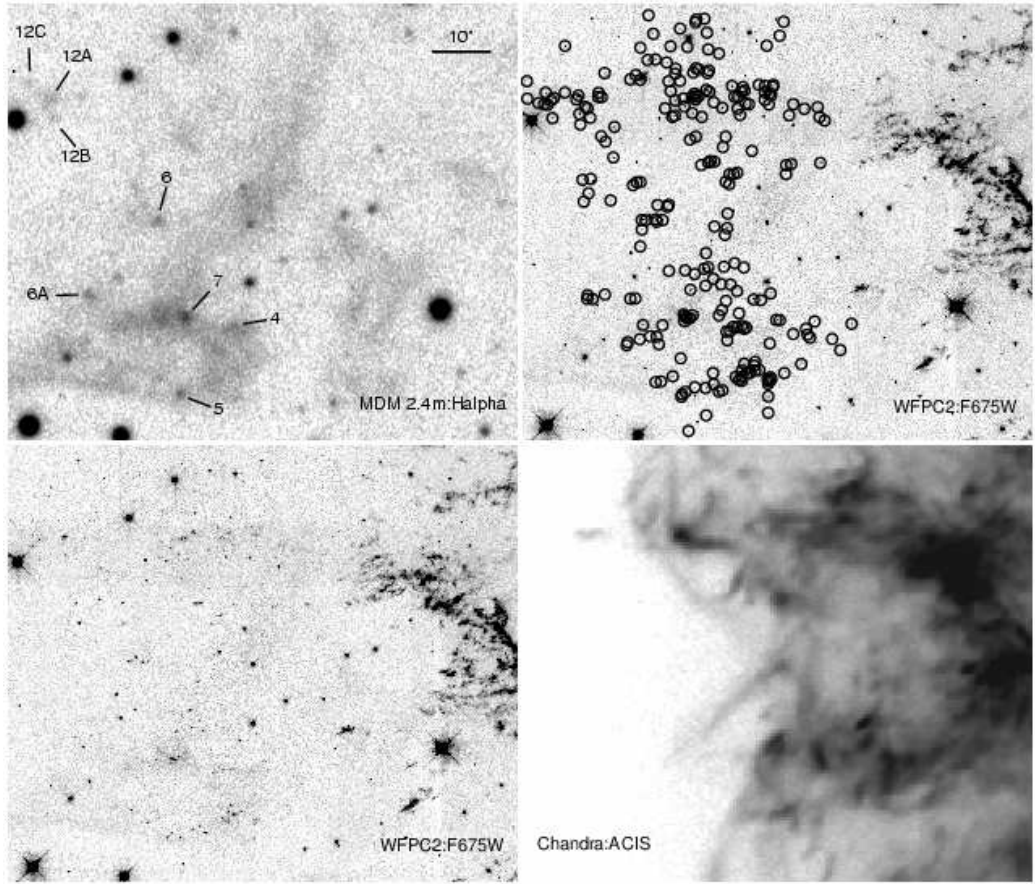} 
\caption{{\it Upper left-hand panel}: Broad H$\alpha$ (FWHM = 90 \AA) ground-based image of the eastern 
periphery of Cas A showing the handful of previously identified outlying emission knots.
{\it Lower left-hand panel}: January 2002 {\sl HST} WFPC2 F675W 
image of the same eastern region but now revealing dozens of 
additional outlying ejecta knots, marked with circles in the {\it upper right-hand panel}. 
{\it Lower right-hand panel}: Same region
showing $0.5 - 10$ keV X-ray emission seen in the 1 Ms exposure {\sl Chandra} image of Cas A. }
\label{fig:MDM_vs_WFPC2} 
\end{figure}

\begin{figure}
\plotone{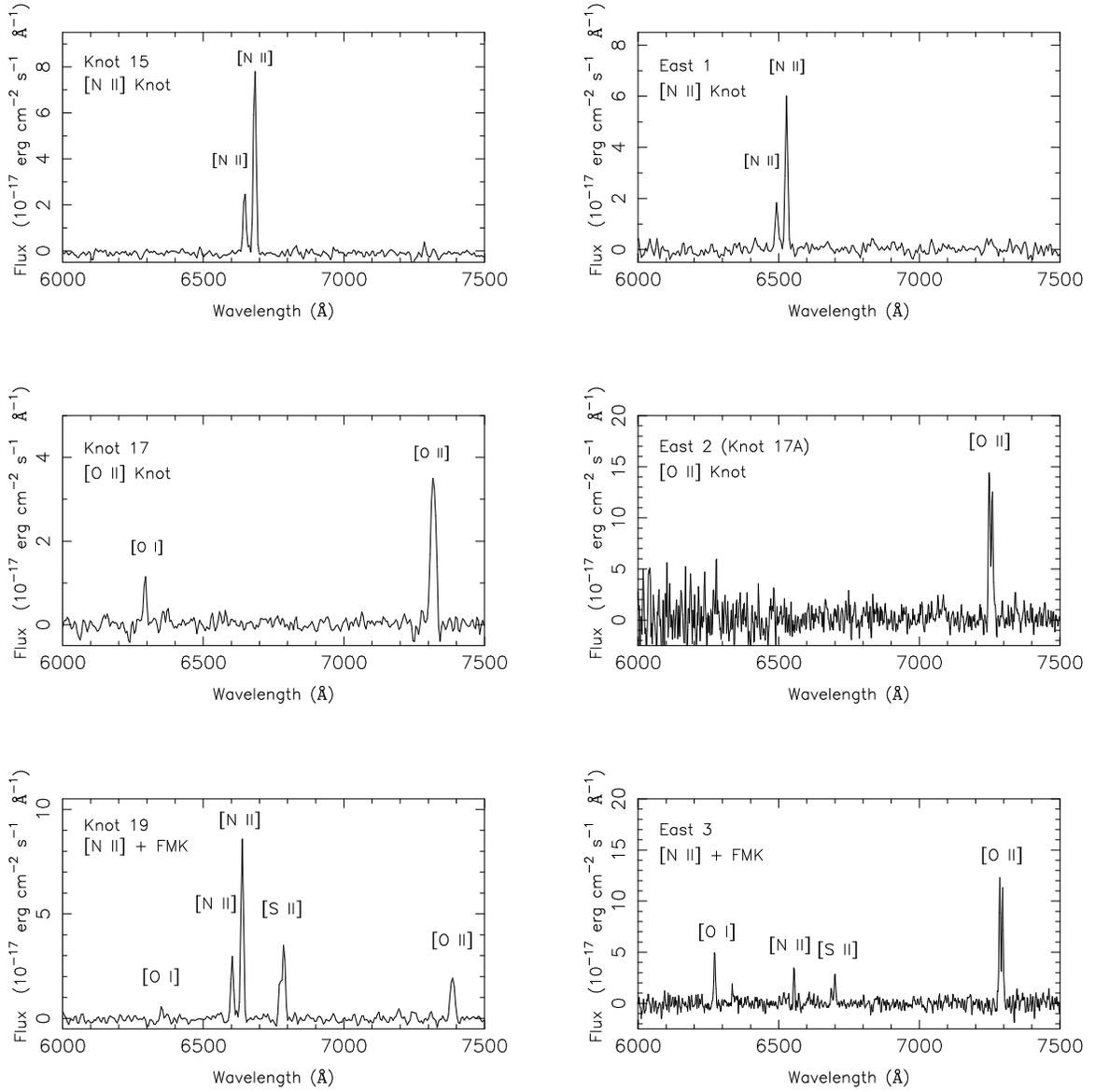}
\caption{Six outer knot spectra  representing the general types of outer knots emission 
([N~II], [O~II], FMK, and mixed FMKs) 
with left-hand panels showing examples of high S/N spectra of knots from various locations around the remnant, 
with the right-hand panels showing spectra of knots located only along the remnant's eastern limb. }
\label{fig:knot_spectra}
\end{figure}

\begin{figure} 
\epsscale{0.90}
\plotone{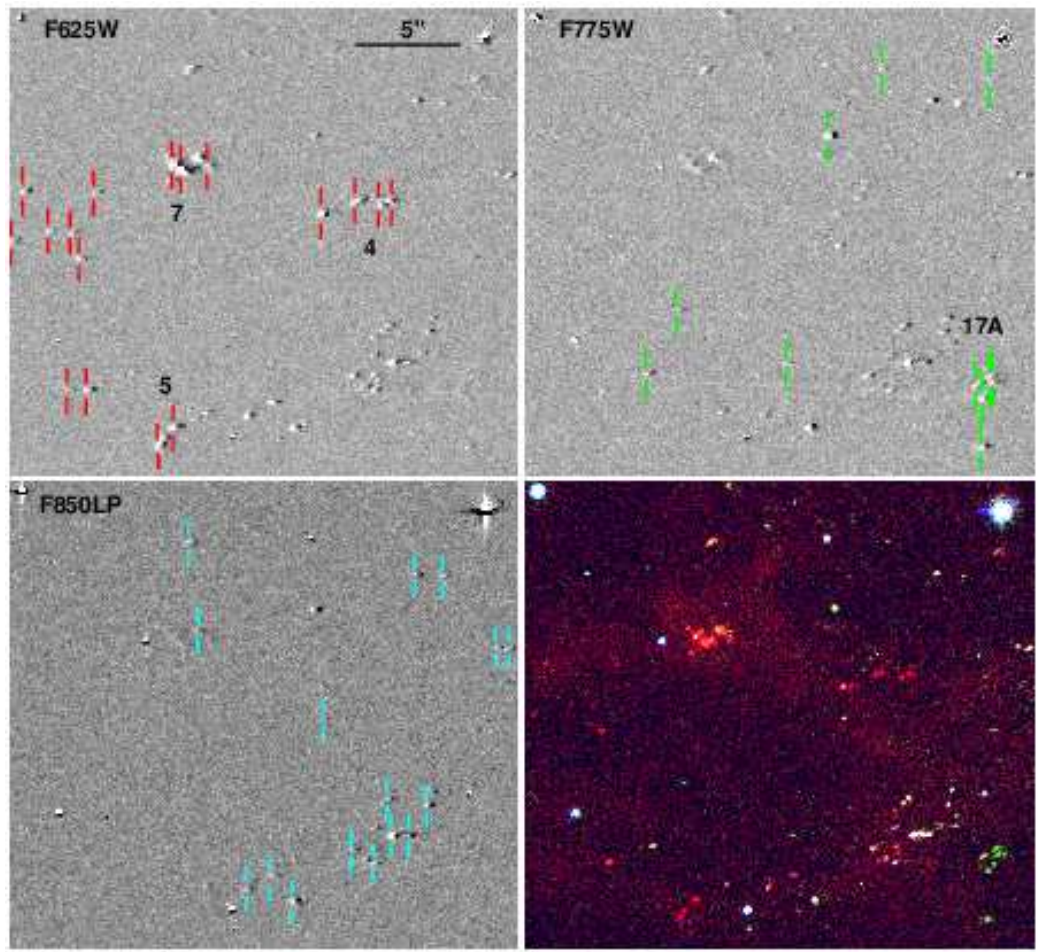} 
\caption{Sections of individual ACS F625W, F775W, and F850LP images are shown for a small portion
of Cas A's eastern limb shown in Figure 1 around outlying ejecta knots 4, 5, and 7, along
with a three color composite image (lower right-hand panel) with the images colored
red, green, and blue respectively. The three greyscale images were produced by subtracting the March from the December
2004 F625W, F775W, and F850LP filter images, while the color composite was made
from just the March 2004 images. Note the group of `green' F775W strong ([O~II]) knots  
in the lower right-hand edge of the color image.}  
\label{fig:individual_acs_images} 
\end{figure}

\begin{figure}
\plotone{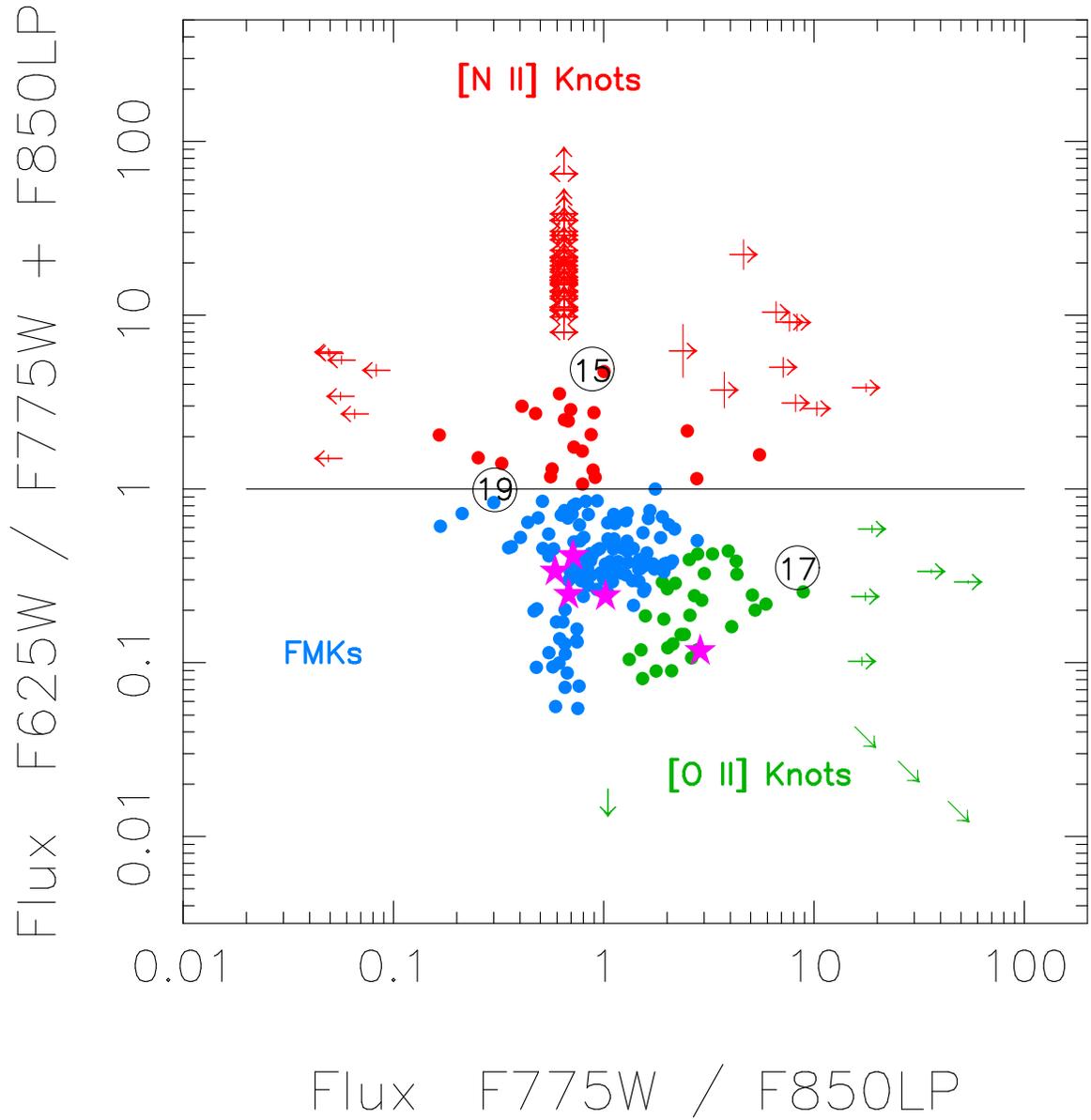}
\caption{Plot of observed {\sl HST} ACS/WFC image fluxes for the eastern outer ejecta knots
for the ratios of F625W/F775W versus F775W/F850LP. 
Knot color coding is  
blue indicating FMK-like knots, red for [N II] emission 
strong knots, and green [O II] emission dominated knots. Arrows indicate a non-detection in at least one filter. 
Lines on the arrow symbols indicate how a detection equal to that of the background flux level in the non-detection filter
would change the knot's plotted position.
The purple star symbols represent SDSS filter ratios for
the five, bright main shell knots listed in Table 2. }
\label{fig:color_by_flux}
\end{figure}

\begin{figure}
\plotone{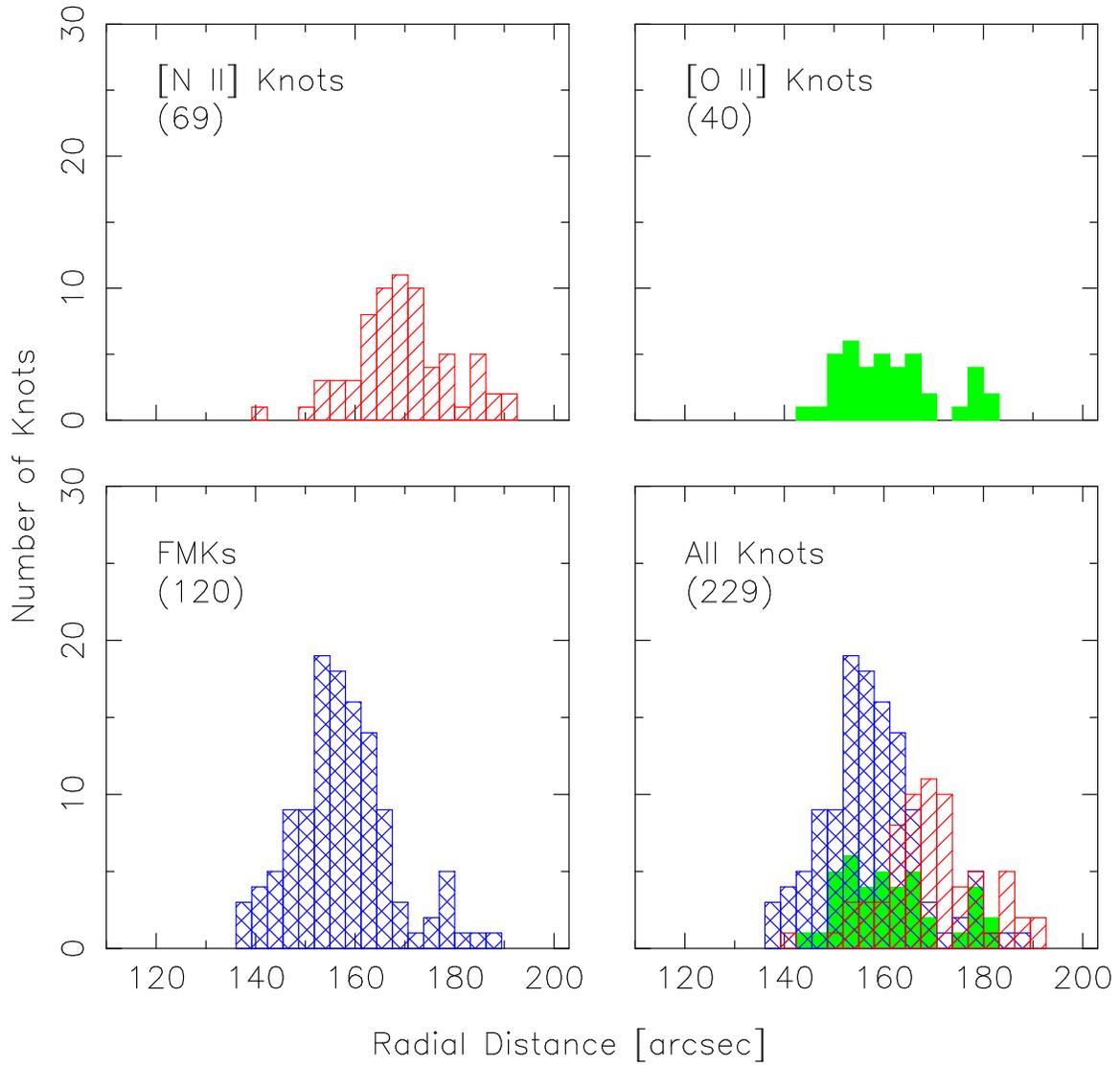}
\caption{Histograms showing the radial distribution of the three types
of outlying ejecta knots located in the eastern region of Cas A.
Numbers in parenthesis indicate the
number of knots in each category.}
\label{fig:radial_dist_SE}
\end{figure}

\begin{figure}
\plotone{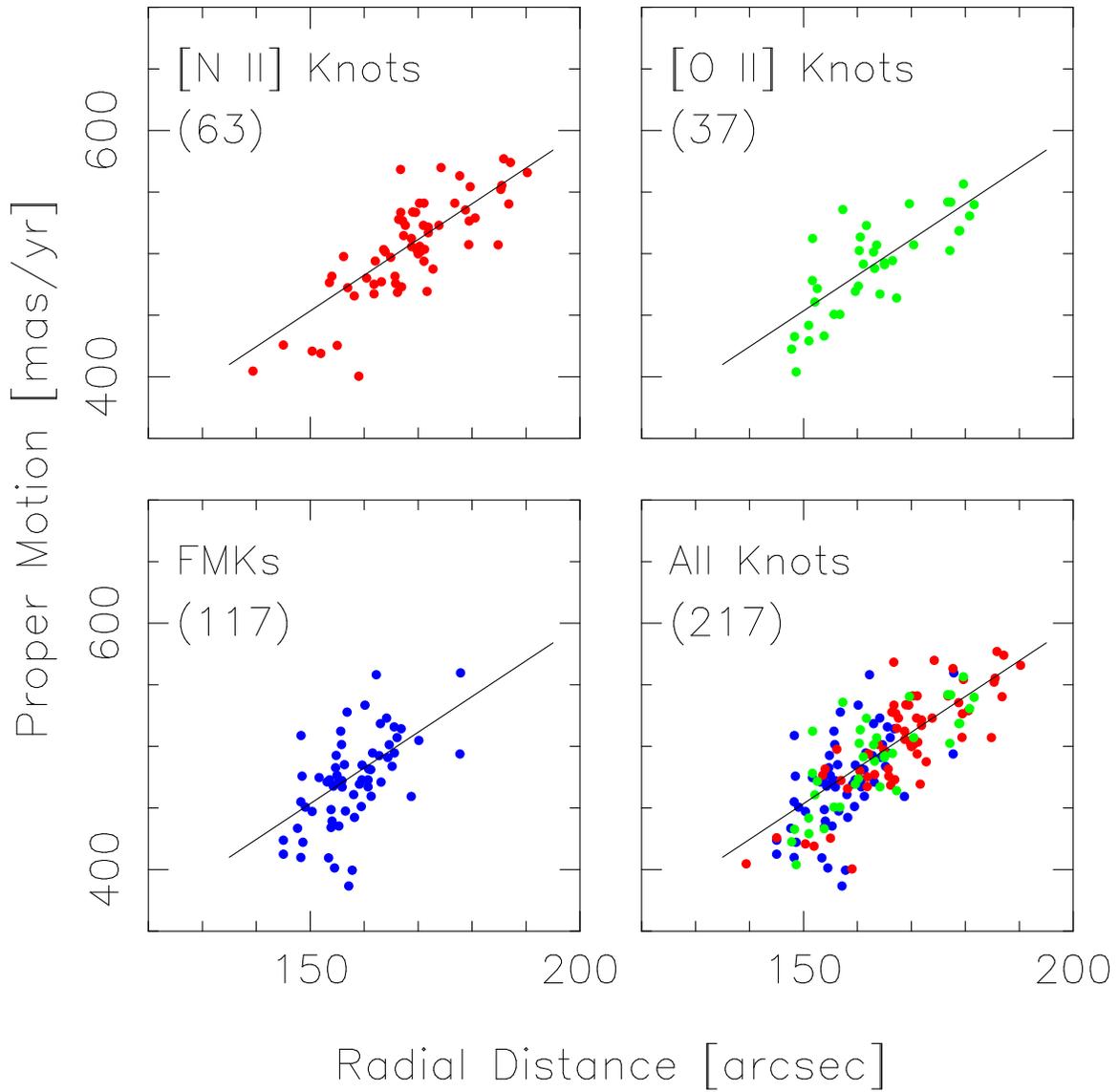}
\caption{Plots showing measured 2000.0 to 2002.0 proper motions
for the three types of outlying ejecta knots located in the eastern region of Cas A.
Numbers in parenthesis indicate the
number of knots in each category.  } 
\label{fig:mu_vs_distance}
\end{figure}

\begin{figure}
\plotone{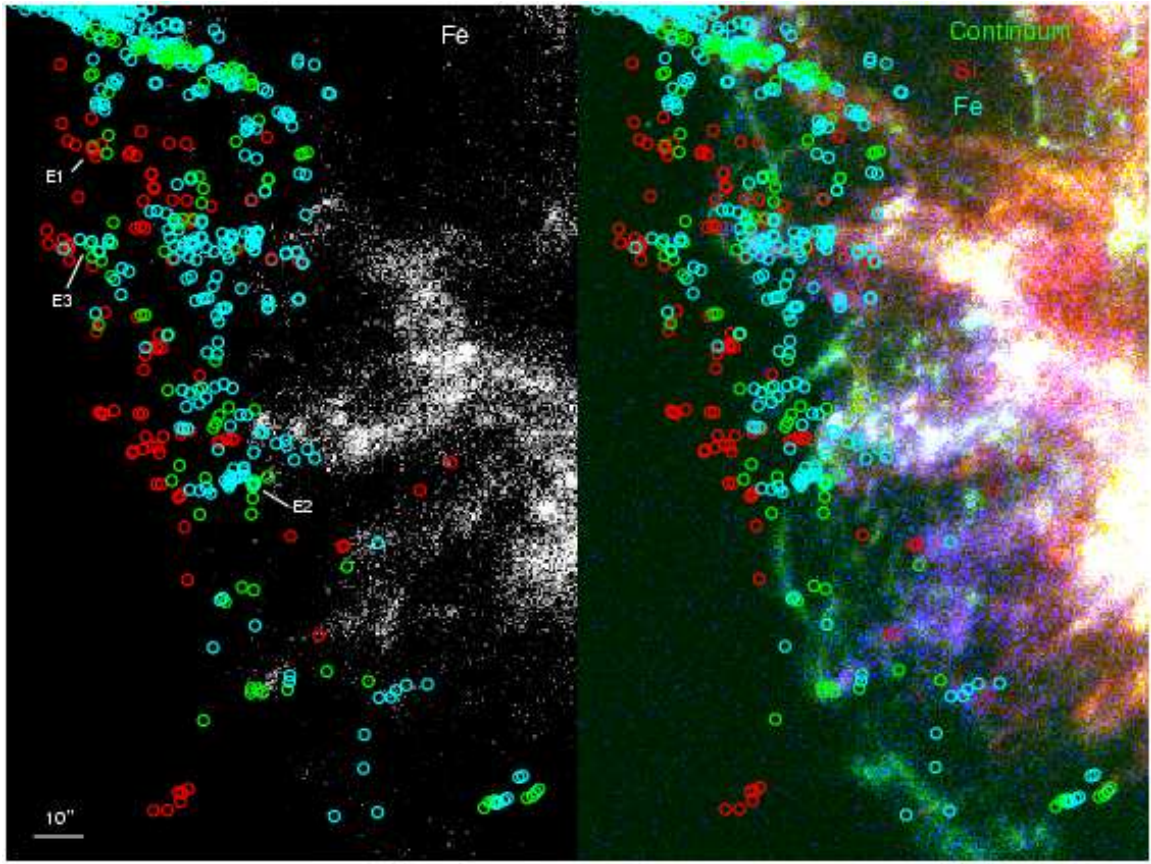}
\caption{Optical outer knots positions along Cas A's eastern limb shown
with respect to color coded {\sl Chandra} 1 Ms X-ray image   
(red = Si He $\alpha$ 1.8--2.0 keV, blue = Fe K 6.5--7.0 keV, green = continuum 4.2--6.3 keV; right panel) and only 
the Fe emission (left panel). Color-coding for the optical knot symbols is the same as in Figure 4. Also marked are locations of three knots with
optical spectra (Knots E1--E3; see Fig.\ 2). }
\label{fig:Chandra_and_optical}
\end{figure}


\begin{thebibliography}{}

\bibitem[Arnett et al.(1989)]{Arnett89} Arnett, D., Fryxell, B., \& Mueller, E.\ 1989, 
         \apjl, 341, L63 
\bibitem[Bertin \& Arnouts(1996)]{Bertin96} Bertin, E., \& 
         Arnouts, S.\ 1996, \aaps, 117, 393 
\bibitem[Blair et al.(2000)]{Blair00} Blair, W.~P., Morse, J.~A., Raymond, J.~C., Kirshner, R.~P.,  
	Hughes, J.~P., Dopita, M.~A., Sutherland, R.~S., Long, K.~S., \& Winkler, P.~F.\
        2000, \apj, 537, 667 
\bibitem[Chevalier \& Kirshner(1978)]{ck78} Chevalier, R. A., \& Kirshner,
          R. P. 1978, \apj, 219, 931
\bibitem[Chevalier \& Kirshner(1979)]{ck79} Chevalier, R. A., \& Kirshner,
         R. P. 1979, \apj, 233, 154 
\bibitem[Delaney \& Rudnick(2004)]{DR04} Delaney, T., \& Rudnick, L.\ 2004,
         Advances in Space Research, 33, 422
\bibitem[Douvion et al.(1999)]{Douvion99} Douvion, T., Lagage, P.~O., \& Cesarsky, C.~J.\ 
         1999, \aap, 352, L111 
\bibitem[Fabian et al.(1980)]{Fabian80} Fabian, A. C., Willingale, R.,
           Pye, J. P., Murray, S. S., \& Fabbiano, G. 1980, \mnras, 193, 175
\bibitem[Fesen(2001)]{Fesen01}Fesen, R. A. 2001, \apjs, 133, 161
\bibitem[Fesen et al.(2001)]{Fesenetal01} Fesen, R.~A., Morse, J.~A., Chevalier, R.~A., 
         Borkowski, K.~J., Gerardy, C.~L., Lawrence, S.~S., \& van den Bergh, S.\ 2001, 
         \aj, 122, 2644 
\bibitem[Fesen \& Becker(1991)]{Fes91} Fesen, R. A., \& Becker, R. H.
          1991, \apj, 371, 621
\bibitem[Fesen, Becker, \& Blair(1987)]{Fes87} Fesen, R. A., Becker, R. H., \& Blair, W. P.
         1987, \apj, 313, 378
\bibitem[Fesen \& Gunderson(1996)]{Fes96} Fesen, R. A., \& Gunderson, K. S.
         1996, \apj, 470, 967
\bibitem[Ford et al.(1998)]{Ford98} Ford, H. C., et al. 1998, Proc. SPIE, 3356, 234
\bibitem[Garc\'{i}a-Segura, Langer, \& Mac Low(1996)]{GS96} Garc\'{i}a-Segura, G.,
          Langer, N., \& Mac Low, M.-M. 1996, \aap, 316, 133
\bibitem[Gotthelf et al.(2001)]{Gotthelf01} Gotthelf, E.~V., Koralesky, B., Rudnick, L., 
          Jones, T.~W., Hwang, U., \& Petre, R.\ 2001, \apjl, 552, L39 
\bibitem[Hammell \& Fesen(2006)]{HF06} Hammell, M. C., \& Fesen, R. A. 2006, in preparation
\bibitem[Hughes et al.(2000)]{Hughes00} Hughes, J. P., Rakowski, C. E.,
         Burrows, D. N., \& Slane, P. O. 2000, \apj, 528, L109
\bibitem[Hurford \& Fesen(1996)]{Hur96} Hurford, A. P., \& Fesen, R. A.
         1996, \apj, 469, 246
\bibitem[Hwang, Holt, \& Petre(2000)]{Hwang00} Hwang, U., Holt, S.~S., \& Petre, R.\ 2000
        \apj, 537, L119
\bibitem[Hwang \& Laming(2003)]{Hwang03} Hwang, U., \& Laming, J.~M.\ 2003, \apj, 597, 362 
\bibitem[Hwang et al.(2004)]{Hwang04} Hwang, U., et al.\ 2004, 
         \apjl, 615, L117 
\bibitem[Kamper \& van den Bergh(1976)]{Kvdb76} Kamper, K. \& van den Bergh, S.
         1976, \apjs, 32, 351
\bibitem[Kifonidis et al.(2000)]{Kifon00} Kifonidis, K., Plewa, T., Janka, H.-T., 
         \& M{\" u}ller, E.\ 2000, \apjl, 531, L123 
\bibitem[Kifonidis et al.(2003)]{Kifon03} Kifonidis, K., Plewa, T., Janka, H.-T.,
         \& M{\" u}ller, E.\ 2003, \aap, 408, 621
\bibitem[Kirshner \& Chevalier(1977)]{KC77} Kirshner, R. ., \& Chevalier, R. A. 1977,
        \apj, 218, 142
\bibitem[Laming \& Hwang(2003)]{Laming03} Laming, J.~M., \& Hwang, U.\ 2003, \apj, 597, 347 
\bibitem[Laming \& Hwang(2005)]{Laming05} Laming, J.~M., \& Hwang, U.\ 2005, \apss, 298, 33 
\bibitem[Langer \& El Eid(1986)]{Langer86} Langer, N., \& El Eid, M. F. 1986, \aap, 167, 265
\bibitem[Lawrence et al.(1995)]{Law95} Lawrence, S. S., MacAlpine, G. M., Uomoto, A.,
           Woodgate, B. E., Brown, L. W., Oliversen, R. J., Lowenthal, J. D., \&
            Liu, C. 1995, \aj, 109, 2635
\bibitem[Massey \& Gronwald(1990)]{Massey90} Massey, P., \& Gronwald, C. 1990, \apj,
          358, 344
\bibitem[Minkowski(1968)]{min68} Minkowski, R. 1968, in
           ``Stars and Stellar Systems'', 7, 623
\bibitem[Morse et al.(2004)]{Morse04} Morse, J.~A., Fesen, R.~A., Chevalier, R.~A., 
        Borkowski, K.~J., Gerardy, C.~L., Lawrence, S.~S., \& van den Bergh, S.\ 2004, 
        \apj, 614, 727 
\bibitem[Pavlovsky et al.(2004)]{Pavlovsky04} Pavlovsky, C., et al. ``ACS Instrument Handbook'', 
          Version 5.0, (Baltimore: STScI) 
\bibitem[Peimbert \& van den Bergh(1971)]{Pvdb71} Peimbert, M., \& van den Bergh, S.
         1971, \apj, 167, 223
\bibitem[Raymond(1979)]{Raymond79} Raymond, J. C.\ 1979, \apjs,  39, 1
\bibitem[Reed et al.(1995)]{Reed95} Reed, J. E., Hester, J. J., Fabian, A. C.,
         \& Winkler, P. F. 1995, \apj, 440, 706
\bibitem[Reynoso \& Goss(2002)]{Reynoso02} Reynoso, E. M., \& Goss, W. M. 2002,
        \apj, 575, 871
\bibitem[Sirianni et al.(2005)]{Sirianni05} Sirianni, M., et al.\ 2005,  preprint (astro-ph/0507614) 
\bibitem[Thorstensen et al.(2001)]{Thor2001}Thorstensen, J. R., Fesen, R. A.,
        \& van den Bergh, S. 2001, \aj, 122, 297
\bibitem[van den Bergh(1971)]{vdb71} van den Bergh, S. 1971, \apj, 165, 457
\bibitem[van den Bergh \& Dodd(1970)]{vdbD70} van den Bergh, S., \& Dodd, W. W. 1970,
          \apj, 162, 485
\bibitem[Vink et al.(1998)]{Vink98} Vink, J., Bloemen, H., Kaastra, J. S.,
          Bleeker, J. A. M. 1998, \aap, 339, 201
\bibitem[Vink(2004)]{Vink04} Vink, J.\ 2004, New Astronomy Review, 48, 61 
\bibitem[Willingale et al.(2002)]{Willingale02} Willingale, R., Bleeker, J.~A.~M., 
         van der Heyden, K.~J., Kaastra, J.~S., \& Vink, J.\ 2002, \aap, 381, 1039 
\bibitem[Willingale et al.(2003)]{Willingale03} Willingale, R., Bleeker, J.~A.~M., 
         van der Heyden, K.~J., \& Kaastra, J.~S.\ 2003, \aap, 398, 1021 
\bibitem[Winkler et al.(1991)]{Winkler91} Winkler, P. F., Roberts, P. F.,
          Kirshner, R. P. 1991, in ``Supernovae: The Tenth Santa Cruz Summer Workshop
           in Astronomy and Astrophysics'', ed. S.E. Woosley (Springer-Verlag: New York), p 652
\bibitem[Woosley et al.(1993)]{Woosley93} Woosley, S.~E., Langer, 
         N., \& Weaver, T.~A.\ 1993, \apj, 411, 823 
\bibitem[Woosley \& Weaver(1995)]{WW95} Woosley, S. E., \& Weaver, T. A. 1995, \apj, 448, 315

\end{thebibliography}
\end{document}